\documentclass [12pt]{article}
\usepackage{graphicx,amssymb,amsmath,bm}
\usepackage[symbol*]{footmisc}
\usepackage[cp1251]{inputenc}
\usepackage[english]{babel}
\usepackage{booktabs}

\makeatletter
\def\@biblabel#1{#1.\hskip-0.3em}
\makeatother

\begin{document}
\def\refname{\normalsize \centering \mdseries \bf References}
\def\abstractname{Abstract}

\begin{center}
{\large \bf Deuteron Stripping on Nuclei at Intermediate Energies}
\end{center}

\begin{center}
\bf\text{V.~I.~Kovalchuk}
\end{center}

\begin{center}
\small
\textit{Department of Physics, Taras Shevchenko National University, Kiev 01033, Ukraine}
\end{center}

\begin{abstract}
A general analytical expression for the double differential cross section of inclusive deuteron stripping
reaction on nuclei at intermediate energies of incident particles was obtained in the diffraction approximation.
Nucleon-nucleus phases were calculated in the framework of Glauber formalism and making use of the
double-folding potential. The exact wave function of deuteron with correct asymptotics at short and long
distances between nucleons was used. The calculated angular dependencies of cross sections are in good
agreement with corresponding experimental data.
\end{abstract}

\bigskip
\begin{center}
\bf{1.~Introduction}
\end{center}
\smallskip

The binding energy of deuteron is low. Therefore, when the latter collides with nuclei, inelastic processes
are the most probable ones: the deuteron breakup in the nuclear Coulomb field (mainly at low deuteron energies)
and the deuteron stripping, when one of deuteron's nucleons is absorbed by the target, whereas the other
is released as a reaction product. In the intermediate energy interval, the stripping reaction is mainly a
result of direct interaction (the capture of deuteron's nucleon by the nucleus), and the differential
cross section of reaction is characterized by a sharp peak at particle emission angles $\Theta\ll1$.
The analysis of the angular and energy distributions of cross sections in the deuteron stripping reaction allows
additional information on the residual nucleus structure and reaction mechanisms to be obtained, being one of
the most important sources of spectroscopic data in nuclear physics.

For the first time, the theory of deuteron stripping at intermediate energies was proposed by R.~Serber~\cite{1}
for transparent and opaque target nuclei, making no allowance for the diffuseness of their surface. Later, the
formalism of inclusive deuteron stripping reaction on nuclei was developed by Akhiezer and Sitenko in work~\cite{2}
on the basis of diffraction nuclear model~\cite{3,4}, and its various aspects were afterwards analyzed and improved
by other authors (see~\cite{5,6} and references therein).

The general formula for the inclusive deuteron stripping cross section~\cite{2} is inconvenient for the analysis
and direct numerical calculations, because it contains a fivefold integral. Therefore, it is usually modified for
practical purposes by introducing additional conditions and restrictions (e.g., the nucleus is opaque and non-diffuse;
the deuteron radius is much smaller than the target one; and so on). However, this integral can be transformed into
a general analytical expression if Gaussian-like functions are used as integrands. Gaussoid functions can be used
here as basis ones for the expansion of both the deuteron wave function (the variational problem) and the profile
functions of arbitrary forms. Notice that a similar trick is widely applied in the variational approach to describe
bound states~\cite{7}, to parametrize the charge densities in the ground state of nuclei~\cite{8,9}, and in scattering
problems~\cite{10}, which makes it possible to calculate the corresponding scattering phases and form factors analytically.

\bigskip
\begin{center}
\bf{2.~Formalism}
\end{center}
\smallskip

Light and medium nuclei were selected as targets, because in this case and in the case of intermediate energies,
the Coulomb interaction can be neglected. The spins of the deuteron's nucleons and the target were also not taken
into account.

The general formula for the differential cross section of deuteron stripping is derived as follows~\cite{2}.
Let a proton be a particle captured by the target nucleus at stripping. The wave function of the neutron released
in this reaction will be presented as a plane wave: ${\psi(\mathbf{r}_{1})=\exp(i\mathbf{k}_{1}\mathbf{r}_{1})}$,
where $\mathbf{k}_{1}$ is the neutron momentum, and $\mathbf{r}_{1}$ its radius vector. The wave functions of the
proton absorbed by the nucleus are coefficients of the integral expansion of deuteron wave function near the nucleus
in series of functions $\psi(\mathbf{r}_{1})$. In other words, the probability amplitude that the neutron
has the momentum $\mathbf{k}_{1}$ and the proton is at the point $\mathbf{r}_{2}$ equals
\begin{equation}
a(\mathbf{k}_{1},\mathbf{r}_{2})=\int d^{(3)}\mathbf{r}_{1}\exp(-i\mathbf{k}_{1}\mathbf{r}_{1})
S_{1}S_{2}\varphi_{0}(\mathbf{r})\psi_{0}(\mathbf{r}_{d}),
\label{eq1}
\end{equation}
where ${S_{i}=1-\omega_{i}}$ are the neutron ($i=1$) and proton ($i=2$) diffraction multipliers, $\omega_{i}$
are the nucleon-nucleus profile functions, and $\varphi_{0}(\mathbf{r})$ and $\psi_{0}(\mathbf{r}_{d})$ the
wave functions of deuteron and its center-of-mass motion, respectively. Let the deuteron move in the positive
direction of $z$-axis. Then, the proton concentration in the $xy$-plane is determined by the squared absolute
value of amplitude (\ref{eq1}),
\begin{equation}
\Big|a(\mathbf{k}_{1},\mathbf{s}_{2})\Big|^{2}=\Big|S_{2}\int d^{(3)}\mathbf{r}_{1}
\exp(-i\mathbf{k}_{1}\mathbf{r}_{1})S_{1}\varphi_{0}(\mathbf{r})\Big|^{2},
\label{eq2}
\end{equation}
where $\mathbf{s}_{2}$ is the impact parameter vector of proton.

Now, integrating the difference between formula (\ref{eq2}) taken at ${S_{2}=1}$ and ${S_{2}\neq1}$, i.\,e. when
the target does not absorb and absorb protons, respectively, over the whole impact plane, we obtain the sought
expression for the double-differential (with respect to the neutron emission angle and energy) cross section,
\begin{equation}
d\sigma_{1}=B(\mathbf{k}_{1})\frac{d\mathbf{k}_{1}}{(2\pi)^{3}},
\label{eq3}
\end{equation}
\begin{equation}
B(\mathbf{k}_{1})=\int d^{(2)}\mathbf{s}_{2}(1-|S_{2}|^{\,2})\Big|\int
d^{(3)}\mathbf{r}_{1}\exp(-i\mathbf{k}_{1}\mathbf{r}_{1})S_{1}\varphi
_{0}(\mathbf{r})\Big|^{2}.
\label{eq4}
\end{equation}
In order to find the angular (energy) distribution of the neutrons arising in the deuteron stripping reaction,
expression (\ref{eq3}) has to be integrated over the longitudinal (transverse) components of vector $\mathbf{k}_{1}$.

As $\varphi_{0}(\mathbf{r})$ in (\ref{eq4}), we use the deuteron wave function (the S-wave) obtained in the
framework of variational method in the Gaussoid basis for the triplet nucleon-nucleon potential from work \cite{11},
\begin{equation}
V(r)=3720.0\exp[-(r/0.488)^{2}]-528.59\exp[-(r/0.976)^{2}],
\label{eq5}
\end{equation}
namely,
\begin{equation}
\varphi_{0}(\mathbf{r})=\sum\limits_{j=1}^{N}{c_{j}}\exp(-d_{j}|\mathbf{r}
_{1}-\mathbf{r}_{2}|^{2}),\quad N=10.
\label{eq6}
\end{equation}
This function has correct asymptotics at short and long distances between nucleons. Besides, it reproduces
the experimental values of deuteron binding energy and deuteron root-mean-square radius~\cite{12} with a high
accuracy.

The nucleon-nucleus profile functions in (\ref{eq4}), which are considered in the framework of Glauber
model~\cite{13},
\begin{equation}
\omega_{i}(s_{i})=1-\exp[-\phi_{i}(s_{i})],
\label{eq7}
\end{equation}
where $\phi_{i}(s_{i})$ is the eikonal phase, can be constructed as follows. Let the distribution of nucleon
density in the impact parameter plane look like
\begin{equation}
\rho_{i}(s_{i})=\rho_{i}(0)\exp(-s_{i}^{2}/a_{N}^{2}),
\label{eq8}
\end{equation}
where $a_{N}^{2}=r_{0}^{2}/{\ln2}$ and $r_{0}^{2}=0.65~\mathrm{fm}^{2}$~\cite{14}. Expanding the density
distribution (experimental~\cite{9} or model) in series of Gaussoid basis functions,
\begin{equation}
\rho_{T}(s)=\sum_{j=1}^{K}\rho_{Tj}\exp(-s^{2}/a_{Tj}^{2}),\quad a_{Tj}
^{2}=R_{rms}^{2}/j\,,
\label{eq9}
\end{equation}
where $R_{rms}$ is the root-mean-square radius of target nucleus, the formula for the eikonal phase
from work~\cite{15} can be generalized:
\begin{equation}
\phi_{i}(s_{i})=\sum_{j=1}^{K}\phi_{ij}(0)\exp\Bigl(-\frac{s_{i}^{2}}
{a_{Tj}^{2}+a_{N}^{2}+r_{0}^{2}}\Bigr),\quad\phi_{ij}(0)=N_{W}\frac{\pi
^{2}\bar{\sigma}_{NN}\rho_{i}(0)a_{N}^{3}\rho_{Tj}\,a_{Tj}^{3}}{a_{Tj}
^{2}+a_{N}^{2}+r_{0}^{2}},
\label{eq10}
\end{equation}
where $N_{W}$ is the normalizing coefficient for the imaginary part of double-folding potential,
and $\bar{\sigma}_{NN}$ the isotopically averaged cross section of nucleon-nucleon interaction.
Substituting (\ref{eq10}) into (\ref{eq7}) and expanding $\omega_{i}$ in series once more, we obtain
\begin{equation}
\omega_{i}(s_{i})=\sum\limits_{j=1}^{K}{\alpha_{ij}}\exp(-s_{i}^{2}/\beta
_{ij}),\quad\beta_{ij}=R_{rms}^{2}/j\,.
\label{eq11}
\end{equation}
Now, substituting functions (\ref{eq6}) and (\ref{eq11}) into (\ref{eq4}) and integrating the result,
we obtain the expression
\begin{equation}
B(\mathbf{k}_{1})=B(\kappa_{1},k_{1z})=\sum\limits_{i=1}^{N}\sum
\limits_{j=1}^{N}{c_{i}}{c_{j}}Y(\lambda^{-1},\kappa_{1},k_{1z}),\quad
\lambda=(d_{i}+d_{j})/2,
\label{eq12}
\end{equation}
where
\begin{equation}
Y(\lambda^{-1},\kappa_{1},k_{1z})=y^{(1)}(\lambda,\kappa_{1},k_{1z}
)-y^{(2)}(\lambda,\kappa_{1},k_{1z}),
\label{eq13}
\end{equation}
\begin{equation}
y^{(1)}(\lambda,\kappa_{1},k_{1z})=4t(\lambda,k_{1z})(y_{11}(\lambda
,\kappa_{1})+y_{12}(\lambda,\kappa_{1})+y_{13}(\lambda,\kappa_{1})),
\label{eq14}
\end{equation}
\begin{equation}
y^{(2)}(\lambda,\kappa_{1},k_{1z})=t(\lambda,k_{1z})(y_{21}(\lambda,\kappa
_{1})+y_{22}(\lambda,\kappa_{1})+y_{23}(\lambda,\kappa_{1})),
\label{eq15}
\end{equation}
\begin{equation}
t(\lambda,k_{1z})={\pi}^{4}{\lambda}^{3}
\exp\Bigl(-\frac{\lambda k_{1z}^{2}}{2}\Bigr),
\label{eq16}
\end{equation}
\begin{equation}
y_{11}(\lambda,\kappa_{1})=\exp\Bigl(-\frac{\lambda\kappa_{1}^{2}}{2}\Bigr)
\sum\limits_{i=1}^{K}\alpha_{2i}\beta_{2i},
\label{eq17}
\end{equation}
\begin{equation}
y_{12}(\lambda,\kappa_{1})=-2\sum\limits_{i=1}^{K}\sum\limits_{j=1}^{K}
\frac{\alpha_{1i}\beta_{1i}\,\alpha_{2j}\beta_{2j}}{\lambda+\beta_{1i}
+\beta_{2j}}\exp\Bigl(-\frac{\lambda+2\beta_{1i}+2\beta_{2j}}
{\lambda+\beta_{1i}+\beta_{2j}}\,\frac{\lambda\kappa_{1}^{2}}{4}\Bigr),
\label{eq18}
\end{equation}
\begin{equation}
y_{13}(\lambda,\kappa_{1})=\sum\limits_{i=1}^{K}
\sum\limits_{j=1}^{K}\sum\limits_{l=1}^{K}\frac{\alpha_{1i}\beta_{1i}\,
\alpha_{1j}\beta_{1j}\,\alpha_{2l}\beta_{2l}}{(\lambda+\beta_{1ij})(\lambda+\beta_{1ij}
+2\beta_{2l})}\exp\Bigl(-\frac{\beta_{1ij}}{\lambda+\beta_{1ij}}
\,\frac{\lambda\kappa_{1}^{2}}{2}\Bigr),
\label{eq19}
\end{equation}
\begin{equation}
y_{21}(\lambda,\kappa_{1})=\exp\Bigl(-\frac{\lambda\kappa_{1}^{2}}{2}
\Bigr)\sum\limits_{i=1}^{K}\sum\limits_{j=1}^{K}\alpha_{2i}\beta_{2i}\,
\beta_{2ij},
\label{eq20}
\end{equation}
\begin{equation}
y_{22}(\lambda,\kappa_{1})=-4\sum\limits_{i=1}^{K}\sum\limits_{j=1}^{K}
\sum\limits_{l=1}^{K}\frac{\alpha_{1i}\beta_{1i}\,\alpha_{2j}\beta_{2j}\,
\beta_{2jl}}{2\lambda+2\beta_{1i}+\beta_{2jl}}\exp\Bigl(-\frac{\lambda+
2\beta_{1i}+\beta_{2jl}}{2\lambda+2\beta_{1i}+\beta_{2jl}}\,
\frac{\lambda\kappa_{1}^{2}}{2}\Bigr),
\label{eq21}
\end{equation}
\begin{equation}
y_{23}(\lambda,\kappa_{1})=\sum\limits_{i=1}^{K}\sum\limits_{j=1}^{K}
\sum\limits_{l=1}^{K}\sum\limits_{n=1}^{K}\frac{a_{1i}\beta_{1i}\,a_{1j}
\beta_{1j}\,a_{2l}\beta_{2l}\,\beta_{2ln}}{(\lambda+\beta_{1ij})
(\lambda+\beta_{1ij}+\beta_{2ln})}\exp\Bigl(-\frac{\beta_{1ij}}{\lambda+\beta_{1ij}
}\,\frac{\lambda\kappa_{1}^{2}}{2}\Bigr),
\label{eq22}
\end{equation}
\begin{equation}
\beta_{ijl}=2\beta_{ij}\beta_{il}/(\beta_{ij}+\beta_{il}),
\quad(i\!=\!1,2;\,\,\,j,l\!=\!\overline{1,K}),
\label{eq23}
\end{equation}
${\mathbf{k}_{1}=}\left\{{\bm{\kappa}_{1},(\mathbf{k}/k)k_{1z}}\right\}$, and $\mathbf{k}$
is the vector of incident deuteron momentum, with $\bm{\kappa}_{1}\,\mathbf{k}=0$.
The components $\kappa_{1}$ and $k_{1z}$ of vector $\mathbf{k}_{1}$ are related
to the neutron energy $T_{1}$ and emission angle $\Theta_{1}$ in the laboratory
reference frame by the formulas~\cite{5}
\begin{equation}
\kappa_{1}=(k/2+k_{1z})\tan\Theta_{1},
\label{eq24}
\end{equation}
\begin{equation}
k_{1z}=\sqrt{m/T}(T_{1}-T/2),
\label{eq25}
\end{equation}
where $m$ is the nucleon mass, and $T$ the initial deuteron energy.

Expressing the components of $d\mathbf{k}_{1}$ in (\ref{eq3}) in the cylindrical coordinates
and using (\ref{eq24}), we obtain the angular distribution of neutrons,
\begin{equation}
\frac{d\sigma_{1}}{d\Omega_{1}}=\frac{1}{(2\pi)^{3}\cos^{3}\Theta_{1}}
\int_{-\infty}^{\infty}(k/2+k_{1z})^{2}B(\kappa_{1},k_{1z})dk_{1z}.
\label{eq26}
\end{equation}
In order to calculate cross section (\ref{eq26}) in the center-of-mass frame, the formulas
of relativistic kinematics from work \cite{16} were used.

\bigskip
\begin{center}
\bf{3.~Results of calculations}
\end{center}
\smallskip

In figure, the neutron angular distributions calculated for the reaction
$^{2}\text{H}(d,n)^{3}\text{He}$ at intermediate energies of incident particles are shown
as an example.
%
\begin{figure}[!h]
\begin{center}
\includegraphics[width=7.0 cm]{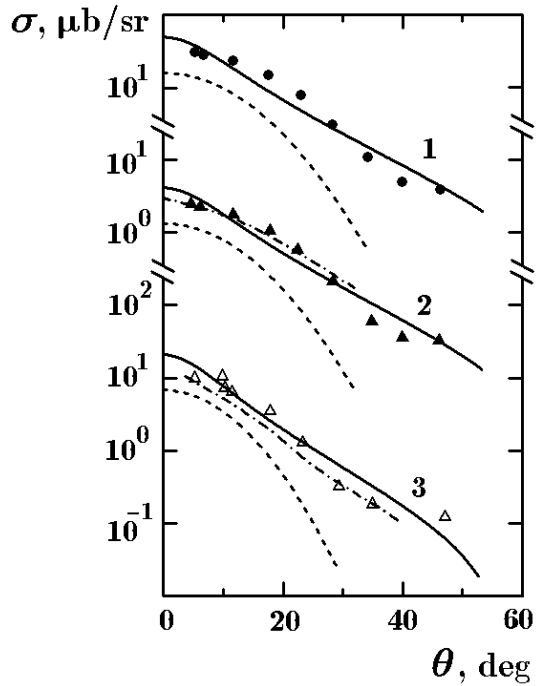}
\end{center}
\caption{Angular distributions of neutrons in the reaction $^{2}$H(d,n)$^{3}$He
at $T=787$ (1), 858 (2), and 1242~MeV (3). See other explanations in the text.
Experimental data were taken from work~\cite{17}.
\label{fig1}}
\end{figure}
%
The solid curves demonstrate the results of cross section calculations with exact deuteron
wave function (\ref{eq6}); the same function was applied while constructing the target density
distribution (\ref{eq9}) with $K=10$. The dashed curves were calculated making use of the model function
\begin{equation}
\varphi_{0}(\mathbf{r})=(2\xi/\pi)^{3/4}\exp(-\xi|\mathbf{r}_{1}
-\mathbf{r}_{2}|^{2}).
\label{eq27}
\end{equation}
Here, the parameter $\xi=0.049$~fm$^{-2}$ was so chosen that (\ref{eq27}) would reproduce
the experimental root-mean-square radius of deuteron~\cite{12}. The dash-dotted curves
reproduce the results of cross section calculations made in work~\cite{17} in the framework
of the virtual pion exchange model. No fitting parameters were used when calculating
cross sections (\ref{eq26}), except for the normalization factor $N_{W}$ for the imaginary
part of double-folding potential in (\ref{eq10}). The relevant $N_{W}$-values were equal
to 0.68 (at $T=787$~MeV), 0.49 (858~MeV), and 0.15 (1242~MeV).

The behavior of calculated curves brings us, first of all, to a conclusion that it is highly
important that the wave function of incident particle with correct asymptotics should be used
in similar calculations. Model function (\ref{eq27}) has a good asymptotic at short internucleon
distances, but the corresponding cross sections decrease more rapidly than experimental values as
the nucleon emission angle $\Theta$ increases (dashed curves). From a comparison between the cross
sections calculated with exact wave function (\ref{eq6}) and the experimental data, it follows that
the behavior of deuteron nucleon density in the tail section of distribution is crucial for
the satisfactory description of experiments (solid curves). Whence a conclusion can be drawn
that the deuteron stripping is a surface reaction~\cite{18}.

\bigskip
\begin{center}
\bf{4.~Conclusions}
\end{center}
\smallskip

The majority of experimental and theoretical works devoted to the researches of deuteron stripping
reactions on nuclei were published in 1960s-1970s. Interest revived recently to this reaction
(see~\cite{6} and references therein) is associated with intensive studies of unstable nuclei.
In this connection, the $(d,N)$ processes may turn out a unique tool for extracting spectroscopic
information. The main result of this work is the exact analytical expression for the corresponding
cross section obtained by transforming integrands in the general formula. Such an approach can also
be used in other similar problems if the relevant integrands can be expanded in series of Gaussoid
basis functions.

Concerning the result of this work, the universal character of its possible application should be
emphasized. The matter is that, in its most general definition~\cite{19}, the inclusive stripping reaction
means that one of the incident particle fragments becomes removed from the particle and participates
in an unobserved interaction subprocess with the target. The subprocess can be arbitrary:
from inelastic scattering to nuclear fusion (really, general expression (\ref{eq4}) contains all
information on the input channel and only partial on the output one). Preliminary calculations show
that double-differential cross section (\ref{eq3}) with $B(\mathbf{k}_{1})$ calculated by formulas
(\ref{eq12})--(\ref{eq23}) successfully describes experimental data for the $(d,pn)$ process, in which
the spectrum of output protons is registered~\cite{20}. In our opinion, the formulas obtained in this
work will also allow one to analyze experimental data on the stripping, pickup, and breakup reactions
for light and heavy ions (provided that the projectile wave function and the corresponding
cluster-nucleus potential are known).

The stripping problem considered above can also be generalized to the case when the spin-orbit
interaction is taken into account. The difference from this work is reduced to the appearance of the
corresponding operator in the expression for profile function. Then, using the density
matrix formalism and carrying out required expansions in the Gaussoid basis, it is possible to derive
an analytical formula for the polarization of particles arising in the stripping reaction.

\end{document}